\documentclass[12pt]{article}

\usepackage{amsfonts,amssymb}

\newcommand{\be}{\begin{equation}}
\newcommand{\ee}{\end{equation}}
\newcommand{\ben}{\begin{displaymath}}
\newcommand{\een}{\end{displaymath}}
\newcommand{\bea}{\begin{eqnarray}}
\newcommand{\eea}{\end{eqnarray}}
\newcommand{\bean}{\begin{eqnarray*}}
\newcommand{\eean}{\end{eqnarray*}}

\def\l {\lambda}
\def\a {\alpha}

\def\b {\beta}

\def\d {\delta}
\def\s {\sigma}

\renewcommand{\O}{\Omega}

\newcommand{\adss}[2]{\mbox{$AdS_{#1}\times {S}^{#2}$}}

\newcommand{\bra}[1]{\mbox{$\langle #1 |$}}
\newcommand{\ket}[1]{\mbox{$| #1 \rangle$}}

\newcommand{\ie}{{\it i.e.}}

\newcommand{\tr}{\mbox{Tr}}

\newcommand{\commentout}[1]{}

\newcommand{\beq}{\begin{equation}}
\newcommand{\eeq}{\end{equation}}
\newcommand{\beqr}{\begin{displaymath}}
\newcommand{\eeqr}{\end{displaymath}}
\newcommand{\beqa}{\begin{eqnarray}}
\newcommand{\eeqa}{\end{eqnarray}}
\newcommand{\beqar}{\begin{eqnarray*}}
\newcommand{\eeqar}{\end{eqnarray*}}
\renewcommand{\k}{\kappa}
\newcommand{\m}{\mu}
\newcommand{\n}{\nu}
\newcommand{\p}{\phi}

\newcommand{\vn}{\ensuremath{\vec{n}}}
\newcommand{\vs}{\ensuremath{\vec{\sigma}}}

\newcommand{\tl}{\ensuremath{\tilde{\lambda}}}

\newcommand{\sql}{\ensuremath{\sqrt{\lambda}}}
\newcommand{\bv}{\ensuremath{\bar{v}}}

\newcommand{\rf}[1]{(\ref{#1})}

\def \ci{\cite}

\def\be{\begin{equation}}
\def\ee{\end{equation}}
\def\ba{\begin{eqnarray}}
\def\ea{\end{eqnarray}}

\def\a{\alpha}
\def\b{\beta}

\def\p{\phi}

\def\P{{\bf P}}

\def\tr{{\rm  tr}}
\def \del{\partial}
\def \a {\alpha}

\def\s{\sigma}

\def\ov{\over}
\def\la{\label}

\def\LL{{\mathcal L }}

\def\b{\beta}
\def\l{\lambda}

\def \De {{\mathcal D}}

\def \adss{$AdS_5 \times S^5$\ }

\def \sql {\sqrt{\lambda} }

\def \p {\phi}

\def \ov {\over}
\def \s{\sigma}

\def \ta{\tau}

\def \ha {{1 \over 2}}

\def \la{\label}

\def \k {\kappa}
\def\foot{\footnote}

\def\det{\hbox{det}}
\def \ci {\cite}

\def \foot {\footnote}
\def \bi{\bibitem}
\def \tr {{\rm tr}}
\def \ha {{1 \over 2}}
 
\def \vv {{\rm v}}  
\def \tl {{\tilde \l}}
\def \XX {{\rm X}}
\def \ta {{\tilde \a}}
\def \fo { {1\ov 4}}
\def \ep {\epsilon}
\def \inti {{\int^{2\pi}_0 {d \sigma \ov 2 \pi}}}
\def \be {\bea}
\def \ee {\eea}
\def \d {\partial}

\def \P {\Phi}
\def \l  {\lambda}
\def \tl {{\tilde \l}}

\def \bv {v^*} 
\def \vv {{\rm v}}
\def \LL {{\mathcal L}}

\def \N {{\mathcal N}} 
\def \S {{\rm S}} 
\def \vn {\vec n}
\def \tl {\td \l} 
\def \td {\tilde} 
\def \Prod {\Pi}\def \O {{\mathcal O}}
\def \Q {{\rm  Q}}
\def \D {\Delta}
\def \N {{\mathcal N}}
\begin{document}
\def \inti { \int^{2\pi}_0 {d\s \ov 2\pi} }

\vskip-1pt
\vskip0.2truecm
\begin{center}
\vskip 0.2truecm {\Large\bf
Semiclassical  strings  in $AdS_5 
\times S^5$ \\ \vskip 0.1truecm
 and  scalar operators in  ${\mathcal N}=4$ SYM theory 
}
\vskip 1.2truecm
{\bf
 A.A. Tseytlin\footnote{Also at Imperial College London
and  Lebedev  Institute, Moscow}\\
\vskip 0.4truecm
 {\it Department of Physics,
The Ohio State University\\
Columbus, OH 43210-1106, USA}\\
 }
\end{center}
\vskip 0.5truecm
\vskip 0.2truecm \noindent\centerline{\bf Abstract}
\vskip .2truecm
 We review recent progress in quantitative checking 
of AdS/CFT duality in the sector of ``semiclassical''
string  states dual to ``long'' scalar operators of 
 ${\mathcal N}=4$  super Yang-Mills 
theory. In particular, we discuss the effective action 
approach, in which the same sigma model type action 
describing coherent states 
is shown to emerge from  the $AdS_5 \times S^5$
string action and  from the   integrable 
 spin chain Hamiltonian representing the SYM 
 dilatation operator.



\section{Introduction}

The $ \N=4$ SYM theory  is a  remarkable  example of 4-d conformal
 field theory. In the planar ($N\to \infty $) limit  it is parametrized by 
the   't Hooft 
coupling $\lambda=g_{_{\rm YM}}^2N$, and the major first  
step towards the solution of this theory  would be to 
determine the spectrum of anomalous dimensions  $\Delta (\l)$ 
of the  primary operators 
built out of  products of local gauge-covariant fields. 
That this may be possible in principle  is suggested by the AdS/CFT 
duality  implying the existence of hidden integrable 
2-d structure corresponding to \adss string sigma model. 

The AdS/CFT duality implies the equality between the $AdS$ 
energies of quantum  closed string states as functions of the effective string 
tension $T= {\sql \ov 2 \pi}$   and quantum numbers like 
$S^5$ angular momenta  $Q=(J,...)$ and dimensions  of the corresponding 
local SYM operators.  To give a quantitative check of the duality 
one would like to understand   how strings
 ``emerge'' from the field theory,
 in particular,  which (local,  single-trace) 
 gauge theory operators \ci{POL} 
 correspond to which ``excited'' string states and 
 how one may   verify
the matching of  their
 dimensions/energies 
beyond the well-understood BPS/supergravity sector.  
We would like  to use the duality as a guide  to deeper understanding 
of the structure of quantum SYM theory.
 In particular, 
 results motivated by comparison to string theory 
may allow one to ``guess'' the general structure  of the SYM 
anomalous dimension matrix 
 and may also  suggest new methods of computing anomalous
 dimensions  in less supersymmetric gauge theories. 

Below we shall review  recent progress  in  checking AdS/CFT
correspondence in a subsector of string/SYM  states
with large quantum numbers. 
Let us start with brief remarks on SYM and string sides
 of the duality.
The SYM  theory contains a gauge field, 6 scalars $\p_m$ 
and 4 Weyl fermions, all in adjoint representation of $SU(N)$. 
It has global conformal and R-symmetry, i.e. is invariant  under
 $SO(2,4) \times SO(6)$. To determine (in planar limit) 
dimensions  of local 
gauge-invariant operators one in general needs to find the anomalous 
dimension matrix to all orders in $\l$ and then to diagonalize it. 
The special case is that of chiral primary or BPS operators (and their descendants)
$\tr ( \p_{\{ m_1 } ... \p_{ m_k\} } )$ whose dimensions are protected, i.e. 
do not depend on $\l$. The problem of finding dimensions 
appears to simplify also in the  case of ``long'' operators 
containing  large number of fields  under the trace. One example 
is provided by ``near-BPS'' operators \ci{bmn} 
  like $\tr ( \P_1^J \P_2^n ...)+ ...$
where  $J \gg n$, and $\P_k = \p_k + i \p_{k+3}$,  \ $k=1,2,3$. 
Below we will consider ``far-from-BPS'' operators 
like  $\tr ( \P_1^{J_1} \P_2^{J_2} ...)+ ...$ where 
$J_1 \sim J_2 \gg 1$. 

The type IIB string action in \adss space  
has the following structure 
\be \la{ss} 
I = -\ha T \int  d \tau \int^{2 \pi}_0  d \s \left(
\del^p Y^\m \del_p Y^\nu \eta_{\m\n} + 
\del^p X^m \del_p X^n \delta_{mn}  + ... \right)
\ , \ee 
where $ Y^\m Y^\nu \eta_{\m\n} =-1, \  \ X^m X^n \delta_{mn}=1 \ , \ 
\eta_{\m\n} = (-++++-)$,  
$T= { \sql \ov 2 \pi} $ and dots stand for the fermionic  terms 
\ci{mt}  that ensure that this model defines a 2-d  conformal field theory.
The closed string states can  be  classified by the values of the Cartan 
charges of the obvious symmetry group $SO(2,4) \times SO(6)$, 
i.e. $(E,S_1,S_2; J_1,J_2,J_3)$, i.e. by the $AdS_5$ energy, two spins in $AdS_5$ and 3 spins in $S^5$.
 The mass shell condition  gives a relation 
$E=E(Q, T)$. Here  $T$ is the string tension and 
$Q=(S_1,S_2,J_1,J_2,J_3; n_k)$ where 
$n_k$  stand for higher conserved charges
 (analogs of oscillation numbers in flat space).

According to AdS/CFT duality  quantum closed string states in \adss 
 should be dual to quantum SYM states at the boundary $R \times S^3$ 
or, via radial quantization, 
 to local single-trace operators at the origin of $R^4$. 
The energy of a string state  should then be equal to the dimension 
of the corresponding SYM operator, 
$E( Q,T)= \D ( Q, \l)$,  where on the SYM side  the charges $Q$ 
characterise the operator. By analogy with 
flat space  and ignoring $\a'$ corrections 
(i.e. assuming $R \to \infty$ or $\a' \to 0$) 
the excited  string states are expected to have energies
$E \sim { 1 \ov \sqrt{\a'}} \sim  \l^{1/4}$  \ci{gkp}  which represents 
a non-trivial prediction for strong-coupling asymptotics of SYM dimensions.
In general, the natural (inverse-tension)  perturbative
expansion on the string side 
will be given by $\sum_n  {c_n \ov (\sql)^n} $, while 
on the SYM side the usual planar perturbation theory  will give 
the eigenvalues of the anomalous dimension matrix as 
$\D= \sum_n a_n \l^n$. The AdS/CFT duality  implies 
that the two expansions are  to be the 
strong-coupling and weak-coupling asymptotics of the same function. 
To check the relation $E= \D$ is then a non-trivial  problem. 
On the symmetry grounds, this  can be shown
 in the case of 1/2  BPS (chiral primary) operators 
dual to supergravity states (``massless'' or ground state string modes)
since their energies/dimensions are protected from corrections.

For generic non-BPS states  the situation looked hopeless until 
the  remarkable suggestion \ci{bmn,gkp} that a progress in
 checking duality
 can be made by concentrating on a subsector of states with large 
(``semiclassical'') values of quantum numbers, $Q \sim T \sim \sql$
(here $Q$ stands for generic quantum number like spin 
in $AdS_5$ or $S^5$ or an oscillation number)
and considering a  new limit 
\be  \la{li}
Q \to \infty \ , \ \ \ \ \ \ \ \ \  
\tl \equiv {\l \ov Q^2} = {\rm fixed}   \ .
\ee 
On the string side ${Q \ov \sql} = { 1 \ov \sqrt{ \tl}}$ 
plays the role of the semiclassical parameter 
(like rotation frequency) which can then be taken to be large. 
The energy of such states  happens to be 
$E= Q + f(Q,\l)$.
The duality implies that such semiclassical string states  
as well as near-by fluctuations  should be dual  to 
``long'' SYM operators with large canonical dimension, 
i.e.  containing large number of fields or derivatives 
under the trace.  In this case the duality map becomes more explicit.

The simplest possibility  is to start with a BPS state that 
carries a large quantum number  and consider  small fluctuations 
 near it \ci{bmn}, 
 i.e. a set of {\it near-BPS}  states   characterised 
by   a large parameter \ci{bmn}.
The only non-trivial example of such BPS state is 
represented  by  a 
point-like string moving along  a geodesic in $S^5$ with a 
large angular
 momentum $Q=J$.
Then $E=J$ and the dual operator is $\tr\ \Phi^J$,  $\Phi= \p_1 + i \p_2$. 
The small (nearly point-like)  
closed strings representing near-by  fluctuations 
  are ultrarelativistic, i.e.  their  kinetic energy 
is much larger  than  their mass.
They 
are dual to SYM operators of the form  $ \tr( \P^J ...)$ 
where 
dots stand for a small number of other fields 
and/or covariant derivatives
(and one needs to sum over different orders of the factors
 to find an eigenstate of the anomalous dimension matrix). The 
energies of the small fluctuations  happen to be \ci{mets,bmn}
$E= J + \sqrt{ 1 + n^2 \tl }\  N_n +  O({1\ov J})$.
One can argue in general \ci{ft1,tse1} and check explicitly \ci{par,cal}
that higher-order quantum string sigma model corrections
are indeed suppressed in the limit \rf{li}, i.e. in the 
 large $J$, fixed $\tl= { \l \ov J^2}\equiv \l'$ limit. 
The  remarkable feature of this expression is 
that  $E$ is analytic in $\tl$, suggesting direct comparison with 
perturbative  SYM expansion in $\l$. Indeed, it can be shown 
that the first two $\tl$ and $\tl^2$  terms in the expansion 
of the square root agree precisely 
with the one \ci{bmn} and two \ci{grom} 
(and also three \ci{bks,beit}) 
loop terms in the anomalous dimensions of the corresponding operators.
There is also a interesting  argument \ci{zan} (for a 2-impurity case)
suggesting how the full  $\sqrt{ 1 + n^2 \tl }$ expression can appear
on the perturbative SYM side. However, the general proof of the consistency 
of the BMN limit on the SYM side (i.e. 
that the usual perturbative expansion 
can be rewritten as an expansion in $\tl $ and $1 \ov J$) 
remains to be given; also, 
 to explain why the string and SYM expressions match one should 
show that the string limit  (first $J\to \infty$, then
  $\tl= { \l \ov J^2}\to 0$)  and  the SYM limit 
(first $\l \to 0$, then $J \to \infty$)  produce 
the same expressions for the dimensions (cf. \ci{kvs,SS,bds}). 

If one moves away from the near-BPS limit  and considers, e.g.,  
a non-supersymmetric  state with a large angular momentum 
$Q=S$ in $AdS_5$
\ci{gkp}, a  direct 
 quantitative check of the duality is no longer possible:
here the classical energy is not analytic in $\l$ and 
quantum corrections are no longer suppressed by powers of $1\ov S$.
However, it is still possible to demonstrate a remarkable qualitative 
agreement between $S$-dependence 
of the string energy and SYM anomalous dimension.
The energy  of a folded closed string rotating  at the center of 
$AdS_5$  which is dual to the twist 2 operators on the SYM side
($\tr ( \Phi^*_k D^S \Phi_k), \ \ D= D_1 + i D_2$ and similar operators 
with spinors and gauge bosons that mix at higher loops \ci{klv,klvo})
has the form (when expanded at large $S$): 
$E= S + f(\l)  \ln S + ...$.  On the string side 
$$f(\l)_{_{\l \gg 1 }} = c_0 \sql + c_1 + {c_2 \ov \sql} + ...\ , $$
where $c_0= { 1 \ov \pi}$ is the classical \ci{gkp} and 
$c_1=  - { 3 \ov  \pi} \ln 2$  is the 1-loop \ci{ft1}  coefficient. 
On the gauge theory side one finds the {\it same} $S$-dependence 
of the anomalous dimension with the 
perturbative expansion of the $\ln S$ coefficient being 
$$f(\l)_{_{\l \ll 1 }} = a_1 \l + a_2 \l^2  + a_3 \l^3  + ...\ , $$ 
where 
$a_1= { 1 \ov 2 \pi^2} $ \ci{gw}, $a_2= - { 1 \ov 96 \pi^2} $ \ci{klv}, 
and $a_3=  { 11 \ov 360 \times 64  \pi^2} $ \ci{klvo}.
Like in the case of the SYM entropy \ci{gkt}, here one expects the existence
 of a smooth interpolating function $f(\l)$ that connects the two 
perturbative expansions (indeed, a simple square root 
formula seems to give a good fit \ci{klv,klvo}).

One could   wonder still if  examples of quantitative agreement 
between string energies and SYM dimensions  observed for near-BPS 
 (BMN) states can be found also for more general non-BPS string states. 
Indeed, it was noticed  already  in \ci{ft1} that a string state 
that carries large spin in $AdS_5$ as well as large spin $J=0$ 
in $S^5$ has, in contrast to the above $J=0$ case, 
 an analytic expansion of its energy  in $\tl= {\l \ov J^2}$, 
just as in the BMN case with  $N_n \sim S$. 
 It was observed  in \ci{ft2} that  semiclassical 
 string states carrying several large spins (with at least one of them
being in $S^5$)  have regular expansion of their energy $E$
in  powers of $\tl$   and it was suggested, 
 by analogy with the  near-BPS  case, that $E$ 
 can be matched with 
perturbative expansion for the SYM dimensions. 

For a classical  rotating closed string solution in $S^5$  one has 
$E= \sql {\mathcal E}(w_i) , \  J_i = \sql w_i$ so that 
$E=E(J_i, \l)$ and the key property is that there is no 
$\sql$ factors in $E$ (in contrast to  the case of a single spin 
in $AdS_5$) 
\be \la{hoi} 
E= J + c_1 { \l \ov J} + c_2 { \l^2 \ov J^3} + ...
= J \ \left[ 1 + c_1 \tl + c_2 \tl^2 + ...\right] \ .\ \ee
Here $J= \sum_{i=1}^3 J_i, \ \tl \equiv {\l \ov J^2}$
and $c_n= c_n({J_i \ov J})$ are functions of
  ratios of the spins which are finite in the limit 
$ J_i \gg 1$, $\tl =$fixed.  The  simplest  example of such a solution is 
provided by a circular string rotating in two orthogonal planes in $S^3$ part of $S^5$
with the two angular momenta being equal $J_1=J_2$  \ci{ft2}:
$\XX_1 \equiv X_1 + i X_2 = \cos (n \s) \ e^{i w \tau} , \ \ 
\XX_2 \equiv X_3 + i X_4 = \sin (n \s) \ e^{i w \tau} ,$
with the global $AdS_5$ time being 
$t= \k \tau$.
\  ($Y_5+ i Y_0 = e^{it}$). 
The conformal 
gauge constraint implies 
$\k^2= w^2 + n^2$ and thus 
$E= \sqrt{ J^2  + n^2 \l} $ or $E= 
J ( 1  + \ha  n^2 \tl - { 1 \ov 8 } n^4 \tl^2 + ...) $, 
where $J= J_1+J_2=2 J_1$. For fixed $J$ the  energy thus has a  regular 
expansion in tension (in contrast to what happens 
in flat space  where  $E= \sqrt{ { 2 \ov \a'} J}$). 
Similar expressions \rf{hoi} are found also for more general multispin 
closed  strings \ci{ft2,ft3,ft4,afrt,art,tse2}. 
In particular,  for a folded string string rotating in one plane of
 $S^5$ and with its center of mass orbiting along big circle in another
 plane \ci{ft4} 
the coefficients $c_n$ are transcendental functions 
(expressed in terms of elliptic integrals). 
More generally, the 3-spin solutions are described by an 
integrable Neumann model \ci{afrt,art}  and  the coefficients 
$c_n$ in the energy are expressed in terms of genus two 
hyperelliptic functions. 

To be able  to compare the classical energy  to the SYM 
dimension one should be sure that string $\a'$  corrections 
are suppressed in the limit $J \to \infty, \ \tl=$fixed. 
Formally, this should be  the case since $\a' \sim { 1 \ov \sql} 
\sim { 1 \ov J \sqrt{ \tl}}$, but, 
what is more important, the $1 \ov J $ corrections are again 
analytic in $\tl$
\ci{ft3},  i.e. the expansion in large $J$ and small $\tl$ 
is well-defined on the string side, 
\be \la{eee}
E= J \left[ 1 + \tl ( c_1 + { d_1 \ov J} + ...) 
+  \tl^2 ( c_2 + { d_2 \ov J} + ...)  + ... \right] \ , \ee
with the classical energy \rf{hoi} being the $J\to \infty$ limit 
of the exact expression. The reason for   this  particular form of the energy  \rf{eee}  can be explained 
as follows \ci{tse1,tse2}: we are computing string sigma model 
loop corrections to the mass of a stationary solitonic solution on a 
2-d cylinder (no IR divergences). This theory is conformal 
(due to the crucial presence of fermionic fluctuations) and thus
does not depend on UV cutoff. 
The relevant fluctuations are massive and their
 masses scale as $w \sim 
{1 \ov \sqrt{\tl}}$. As a result,  the inverse mass expansion 
is well-defined and the quantum corrections should be proportional 
to positive powers of $\tl$.

Similar expressions are found  for the energies of small
fluctuations near a given classical solution: as in the BMN case, 
the fluctuation energies are suppressed by extra  factor of $J$, 
i.e. $\delta E = \tl ( k_1 + { m_1 \ov J} + ...) 
+  \tl^2 ( k_2 + { m_2 \ov J} + ...)  + ...$. 

Assuming that the same limit is well-defined also on the SYM side, 
one should then be  able to compare the coefficients in \rf{eee}
to the coefficients  in the anomalous 
dimensions of the corresponding SYM operators \ci{ft2} 
$\tr( \P_1^{J_1} \P_2^{J_2}  \P_3^{J_3}  ) + ...$.
(and also do similar matching for near-by fluctuation modes). 
In practice, it is known at least in principle 
how to compute the dimensions in a different limit: 
first expanding in $\l$  and then expanding in $1 \ov J$.
One may expect that this expansion  of anomalous dimensions 
takes the form equivalent to \rf{eee}, i.e.
\be 
\D= J + \l ( { a_1 \ov J} + { b_1 \ov J^2} + ...) 
      + \l^2  ( { a_2 \ov J^3} + { b_2 \ov J^4} + ...) + ...\ , \ee 
and that  the respective coefficients in $E$
and $\D$  agree with each other. 
The subsequent work \ci{mz1,bmsz,bfst,as,Min2,kru,SS,kmmz,krt,zarL}
did  verify this  structure of $\D$ and moreover 
established the general agreement 
between the two leading coefficients $c_1,c_2$  in $E$ \rf{eee}
and the ``one-loop'' and ``two-loop'' coefficients 
$a_1,a_2$ in $\D$.

To compute $\D$  one is first  to  diagonalize 
anomalous dimension matrix defined on a set of long scalar 
operators. The crucial step was made in \ci{mz1} where it was 
observed 
that the one-loop planar  dilatation operator in the scalar sector 
can be interpreted as a Hamiltonian of an integrable $SO(6)$
 spin chain and thus can be diagonalized even for large 
length $L=J$  by the  Bethe ansatz method.
In the simplest case of (closed)  ``$SU(2)$'' sector  of operators 
$\tr( \P_1^{J_1} \P_2^{J_2}) + ...$  built out of 
two chiral scalars the dilatation operator   can be interpreted as ``spin up'' and 
``spin down'' states of periodic XXX$_{1/2}$  spin chain with length 
$L=J=J_1+J_2$. Then 
the 1-loop dilatation operator becomes equivalent 
to the Hamiltonian of the ferromagnetic Heisenberg model 
\be \la{ferr}
D_1 = { \l \ov (4 \pi)^2} \sum^J_{l=1}
(1 - {\vs}_l \cdot { \vs}_{l+1})  \ . \ee
By considering the thermodynamic 
limit ($J \to \infty$) of the Bethe ansatz 
the 
proposal of  \cite{ft2} was confirmed at the leading order 
of expansion in $\tl$  \cite{bmsz,bfst}: 
for eigen-operators  with $J_1\sim J_2 \gg 1 $ 
it was shown  that 
$\D-J=  \l {a_1 \ov J} + ...$ and a  
 remarkable agreement 
was found between $a_1({J_1 \ov J_2})$ and the coefficient $c_1$
 in the energies of  various 2-spin  string solutions.
As in the BMN  case, it was possible also to match 
the energies of fluctuations 
near the circular $J_1=J_2$ solution 
 with the corresponding eigenvalues 
of \rf{ferr} \ci{bmsz}. 

Similar leading-order agreement between string energies and 
SYM dimensions was observed also in other sectors of states
with large quantum numbers: (i) for  specific 
solutions \ci{ft2,afrt,art} in  the $SU(3)$ sector 
 with 3 spins in $S^5$  dual to $\tr( \P_1^{J_1}
 \P_2^{J_2}  \P_3^{J_3}  ) + ...$ operators 
\ci{Min2,char}; 
(ii) for a folded string state \ci{ft1} 
belonging to  the $SL(2)$ \ci{BS} sector 
with one spin in $AdS_5$ and one spin in $S^5$
 (with $E=
J + S + { \l \ov J} c_1 ( {S\ov J}) + ...$ \ci{ft1,ft2}) 
dual to 
$\tr ( D^S \P^J) + ... $ \ci{bfst};
(iii) in a ``subsector'' of $SO(6)$ states 
containing pulsating (and rotating)  solutions \ci{bmsz,Min2}
which again have regular energy expansion  in the limit 
of large oscillation  number, e.g., 
 $E=L + c_1 { \l \ov L}  + ...$ \ci{Min1}.

\section{Effective actions for coherent states}

The observed agreement between energies of particular 
semiclassical string states and dimensions of the corresponding 
``long'' SYM operators leaves many questions, in particular:
(i) How to understand this agreement  beyond specific examples, 
i.e. in a universal way? 
(ii) Which is the  precise relation between profiles of string 
solutions and the structure of the dual SYM operators?
(iii) How to characterise the set of semiclassical string states 
and dual SYM operators for which the correspondence should work?
(iv) Why agreement works, i.e. why the two limits 
(first $J\to \infty$, and then $\tl \to 0$, or vice versa) 
taken on the string and SYM sides 
give equivalent results?  Should it work to all orders 
in expansion in $\tl$ (and $1\ov J$)? 
 The questions (i),(ii) were addressed 
in \ci{kru,krt,lopez,ST,kt};
an alternative approach based on matching 
the general solution (and integrable structure) 
of the  string sigma model with that of the thermodynamic limit 
of the Bethe ansatz was developed in \ci{kmmz}.
The question (iii) was addressed  in 
\ci{mateos,mik,mikk,kt}, and  the question (iv) -- in 
\ci{SS,bds,afs}.

One  key idea of \ci{kru} (elaborated further 
in \ci{krt,kt})   was that instead of comparing particular 
solutions one should try to match  effective sigma models 
which appear on the string side  and the SYM side.
Another related   idea of \ci{kru,krt,kt} 
was that since ``semiclassical'' string states 
carrying  large quantum numbers are represented in the 
quantum theory  by coherent states, one should be comparing coherent 
string states to coherent SYM states (i.e. to coherent 
states of the spin chain).
Because of the ferromagnetic 
nature of the dilatation operator \rf{ferr} in   the thermodynamic 
limit 
$J=J_1+J_2  \to \infty$ with fixed number of impurities $J_1 \ov J_2$ 
it is  favorable to form large clusters of spins and thus a 
``low-energy''  
approximation and continuum limit  apply, leading to an effective 
``non-relativistic''  sigma model for a coherent-state expectation 
value of the spin operator. 
Taking  the ``large energy''  limit directly 
in the string action gives  a reduced ``non-relativistic'' 
sigma model that describes in a universal way 
the  leading-order $O(\tl)$ corrections 
to the energies of all string solutions in the two-spin sector.
The resulting action 
agrees  exactly \ci{kru}  with   the semiclassical 
 coherent state action describing the 
$SU(2)$ sector of the  spin chain
in the $J \to \infty, \ \tl=$fixed  limit. 
This demonstrates 
how a string action can directly emerge from a gauge 
theory in the large-$N$ limit and 
provides a direct map between  
``coherent''  SYM  states (or operators built out of two holomorphic  
scalars) and all two-spin classical string states.
Furthermore, the correspondence  established at the level
of the action implies also  the matching of fluctuations 
around particular  solutions
 (as in the BMN case) 
and thus  goes  beyond the special examples of 
 rigidly rotating strings. 


Let us briefly 
review the definition of coherent states
(see, e.g.,  \ci{zhang}). For a harmonic oscillator  
($ [a,a^\dagger]=1$) one can define  the coherent state 
as $\ket{u}$ as
$a \ket{u} = u \ket{u}$, where $u$  is a complex number. 
Equivalently, $\ket{u}= R(u) \ket{0}$, where 
$R= e^{ u a^\dagger - u^* a} $ so that acting on $\ket{0}$
$R$ is simply proportional to $e^{ u a^\dagger} $.
Note that $\ket{u}$  can be written as a  superposition 
of eigenstates $\ket{n}$ of the harmonic oscillator Hamiltonian, 
$\ket{u} \sim \sum^\infty_{n=0} { u^n \ov \sqrt{ n !}}
\ket{n}$.
An alternative definition of coherent state is  that it is a state 
with minimal uncertainty for both coordinate 
$\hat q = { 1 \ov \sqrt 2} ( a +  a^\dagger) $ and 
momentum $\hat p = -{ i \ov \sqrt 2} ( a - a^\dagger) $
operators, $\Delta \hat p^2 = \Delta \hat q^2 = \ha$, 
\ $ \Delta \hat p^2 \equiv \bra{u } \hat p^2  \ket{u}
- ( \bra{u } \hat p \ket{u})^2 $. 
For that reason this is the best approximation to a classical state.
If one defines a time-dependent state $\ket{u(t)}
= e^{- i H t} \ket{u}$ then the expectation values of 
$\hat q$ and $\hat p$ 
$\bra{u } \hat q \ket{u} = { 1 \ov \sqrt 2} ( u + u^*), \ 
$ $ \bra{u } \hat p \ket{u} =- { i \ov \sqrt 2} ( u -u^* )$, 
will follow the classical trajectories.
 Starting with the $SU(2)$ algebra 
$[S_3, S_\pm ] = \pm S_\pm, \ [S_+, S_-] = 2 S_3$ 
and considering the $s=1/2$ representation where
 $\vec S = \ha \vec \s$ 
one can define spin coherent state
 as a 
linear superposition of spin up and spin down
states: $\ket{u} = R(u) \ket{0}, $ where 
$ R= e^{ u S_+ - u^* S_-}$,  $ \ket{0} = \ket{\ha, \ha}$
and $u$ is a complex number.
An equivalent way to 
label  the  coherent state is  by a unit 3-vector $ \vec n$ 
defining a point of $S^2$. Then $ \ket{\vec n }
= R( \vec n) \ket{0 }$ where $ \ket{0 } $ 
corresponds to a 3-vector $(0,0,1)$ along the 3rd axis
($ \vec n= U^\dagger \vec \s U, \ U= (u_1,u_2)$)
and $R( \vec n)$ is an $SO(3)$ rotation  from a north pole to 
a generic point of $S^2$. The key property of the coherent state is 
that $\vec n$ determines the expectation value of
 the spin operator:
$\bra{\vec n  } \vec S  \ket{\vec n} = \ha  \vec n$. 

In general, one can rewrite the usual phase space path integral 
as an integral over the overcomplete set of coherent states 
(for the harmonic oscillator this is 
simply the change of variables $ u = { 1 \ov \sqrt 2} ( q + i p)$):
\be Z= \int [du] \ e^{i \S[u]}  \ , \ \ \ \ \ \ \ \ \ 
\S= \int dt \bigg( \bra{u }  i  { d \ov dt} \ket{u}  - 
\bra{u }  H  \ket{u} \bigg) \ , \ee
 where the first (WZ or ``Berry phase'') 
term is the analog of the usual $ p \dot q$ term
 in the phase-space action. Applying this to the case of the 
Heisenberg 
spin chain Hamiltonian 
\rf{ferr} one ends up with  with the following action 
for the coherent state variables $\vec n_l(t)$ at sites $l=1,...,J$
 (see also \ci{fra}):
\be \S=\int dt \sum^J_{l=1}  \bigg[ \vec C (n_l) \cdot \vec n_l  
- { \l \ov 2 (4 \pi)^2 }  ( \vec n_{l+1} 
- \vec n_l)^2  \bigg] \ . \ee
Here $d C= \epsilon^{ijk} n_i d n_j \wedge d n_k$
(i.e. $\vec C$ is a monopole potential on $S^2$). 
In local coordinates (at  each $l$) 
$\vec n = (\sin \theta \ \cos \p, \ 
 \sin \theta \ \sin \p,\ \cos \theta)$, 
\ $\vec  C \cdot d \vec n = \ha \cos \theta  d \p$.
In the limit $J\to \infty$  with fixed $\tl= {\l \ov J^2}$ 
 one can take  a continuum limit 
by introducing the 2-d field $ \vec n(t,\s)= \{ \vec n (t,
 { 2 \pi \ov J} l) \}$. Then 
\be \la{con}
\S= J \int dt \inti  \left[ \vec C \cdot  \del_t  \vec n - 
{1 \ov 8} \tl (\del_\s \vec n)^2  + ... \right] \ , \ee
where dots stand for higher derivative terms suppressed by 
$1 \ov J$. In the limit $J \to \infty$ we are interested in  
 all quantum corrections are thus  suppressed 
by $1 \ov J$, and thus the above action can be treated classically.
The corresponding equation of motion 
$ \dot n_i = \ha \tl \epsilon_{ijk} n_j n''_k$ are the 
Landau-Lifshitz equation
for a classical ferromagnet. 

The action \rf{con} should be describing the coherent states 
of the Heisenberg spin chain in the above thermodynamic limit. 
One may wonder how a similar ``non-relativistic'' 
action may appear  on the string side where one starts 
with a usual sigma model \rf{ss}. 
 To obtain such an  effective action 
 one  is  to 
perform the following procedure \ci{kru,krt,kt}:
(i) isolate a  ``fast'' coordinate $\a$ whose momentum  $p_\a$ is
large in our limit; (ii) gauge fix $t \sim \tau$ and 
$ p_\a \sim J$ (or $\td \a \sim \s$ where $\td \a $ is ``T-dual''
to $\a$); (iii) expand the action in  derivatives of 
``slow'' or `transverse'' coordinates 
(to be identified with $\vec n$).
Let us consider  the $SU(2)$ sector of string states 
carrying two large spins in $S^5$, with string motions restricted 
to $S^3$ part of $S^5$. The relevant part of the \adss metric is then 
$ds^2= - dt^2 + d\XX_i d\XX_i^*$, with $\XX_i \XX_i^*=1$.
Let us 
set $$\XX_1 = X_1 + i X_2 = u_1 e^{i \a}\ , \ \ \ \ \ \ \ \ \ 
\XX_2 = X_3 + i X_4 = u_2 e^{i \a}\ , \ \ \ \ \ \ \ \ 
 u_i u^*_i=1\ . $$
Here $\a$ will be a collective coordinate  associated to the 
total (large) spin in the two planes (which in general 
will be the sum of orbital  and internal spin);  
$u_i$   (defined modulo $U(1)$ 
gauge transformation) will be the  ``slow'' 
coordinates determining the
 ``transverse'' string profile. Then 
$$d\XX_i d\XX_i^* = ( d \a + C)^2 + D u_i Du^*_i  \ , \ \ \ \ \ \  
C= - i u^*_i du_i\ , \ \ \ \ \ \ \ Du_i = d u_i -i C u_i \ ,  $$
 and the second term 
represent the metric of $CP^1$ (this parametrisation corresponds 
to Hopf fibration $S^3 \sim S^1 \times S^2$). 
Introducing $\vec n = U^\dagger \vec \s U, \ U=(u_1,u_2)$ 
we get $d\XX_i d\XX_i^* = (D \a)^2 +  { 1 \ov 4} ( d \vec n)^2 
$, \ $D\a= d \a + C(n)$.
 Writing the resulting sigma model action in phase space form 
and imposing the (non-conformal) 
gauge $t= \tau, \ p_\a =$const$=J$ 
one gets the action \rf{con} with the WZ term $
\vec C \cdot  \del_t  \vec n$
originating from the  $p_\a D \a $ term in the phase-space 
Lagrangian (cf. its origin on the spin chain  side
as an analog of the  `$p \dot q$' in the coherent state 
path integral action). 
Equivalent approach is based on first applying a 2-d 
duality (or ``T-duality'')  $\a\to \td \a $  and then choosing the 
``static'' gauge $t= \tau, \ \td \a ={ 1 \ov \sqrt{ \td \l}  } \s, \
\   { 1 \ov \sqrt{ \td \l}  }= { J \ov \sql}$.  
Indeed, 
starting with \be
\LL =  - \ha \sqrt{ - g} g^{pq}
\big( - \d_p t \d_q t + D_p \a  D_q \a  + D_p u^*_i 
D_q u_i
\big) \ee and 
applying T-duality in $\a$ we get 
\be\LL =  - \ha \sqrt{ - g} g^{pq}
\big( - \d_p t \d_q t + \del_p \ta  \del_q \ta  + D_p u^*_i
 D_q u_i
\big)  +   \ep^{pq} C_p \del_q \ta \ . \ee
Thus the ``T-dual''  background  has no off-diagonal
 metric component 
but has a  non-trivial NS-NS 2-form  coupling 
in the $(\ta,u_i)$  sector.  
It is important that we do not use
conformal gauge here. 
Eliminating the  2-d metric $g^{pq}$ we then get 
the Nambu-type action   
\be \la{namb} 
\LL  =     \ep^{pq} C_p \del_q \ta\  - \sqrt{  h } 
\   ,  \ee
where 
$
h =|\det \  h_{pq}| $ and 
$ h_{pq} = 
- \d_p t \d_q t + \del_p \ta  \del_q \ta 
 + D_{(p}  u^*_i D_{q)} u_i$. 
If we now   fix the  static gauge 
$
t = \tau , \  \ta=  {1 \ov  \sqrt{ \td \l}  } \s $
we finish with the following action 
$I =  J \int dt \inti \ \LL$, where 
\be 
\la{eaa}
\LL  =  C_0  -  \sqrt{  (1+  \tl  |D_1 u_i|^2)( 1 -  |D_0 u_i|^2  )
+  \fo \tl   (D_0 u^*_i D_1 u_i + c.c.)   ^2 }\ . 
\ee
To leading order in $\tl$  this gives  
\be \la{coon}
\LL = - i u^*_i \del_0 u_i - \ha \tl |D_1 u_i|^2 \ , \ee
        which is the same as the $CP^1$ Landau-Lifshitz 
action  
 \rf{con} when written in terms of 
$\vec n$. 
Thus the string-theory  
counterpart   of the WZ term   in the spin-chain
coherent state  effective 
 action  comes from the 2-d  NS-NS WZ term upon
the  static gauge fixing in the ``T-dual''  $\ta$ action \ci{kt}. 

To summarize: 
(i) $(t, \ta)$ are the  ``longitudinal'' coordinates 
that are gauge-fixed (with $\ta$ playing 
the role of string direction or spin chain direction on the
SYM side); (ii)  $U=(u_1,u_2)$  or $\vec n = U^\dagger
\vec \s U$ are ``transverse'' coordinates  that determine 
the semiclassical string profile  and also 
the structure of the coherent operator  on the  SYM side, 
$\tr\ \Prod_\s (u_i \Phi_i) $.
The agreement between the low-energy actions on the spin 
chain and the string side explains not only 
the matching of  
energies of  coherent states for configurations with two
large spins (and near-by fluctuations) but also the matching
of integrable structures observed on specific 
examples in \ci{as,Min2}. 

This leading-order agreement in $SU(2)$ sector has several
generalizations. First, we may include higher-order terms 
on the string side. Expanding 
\rf{eaa} in $\td \l$ and eliminating higher powers of time
derivatives  by field redefinitions (note that 
leading-order equation of motion is 1-st order in time
derivative)
we end up with \ci{krt}
$$
\LL= \vec C \cdot  \dot{  \vec n }- 
{\tl \ov 8}   \vec n'^2  
+ { \tl^2 \ov 32}  ( \vn''^2 - { 3 \ov 4} \vn'^4) 
- { \tl^3 \ov 64}  \big[ \vn'''^2 - { 7 \ov 4} \vn'^2 \vn''^2 
- {25 \ov 2} (\vn'  \vn'')^2 + { 13 \ov 16} \vn'^6\big] 
+ ...   $$
The same $\tl^2$ term is obtained \ci{krt} 
in the coherent state 
action on the spin chain side by starting with the 
sum of the 1-loop dilatation operator \rf{ferr} 
 and the   2-loop term found in   \ci{bks}
\be \la{two}
D_2 = { \l^2 \ov (4 \pi)^4} \sum_{l=1}^J ( \Q_{l,l+2} - 4
\Q_{l,l+1}) \ , \ \ \ \ \ \ \ \ 
\Q_{k,l} \equiv I - \vec\s_k \cdot \vec\s_l \ . \ee 
This explains  the matching of energies and dimensions 
to the first two orders, as  first observed on specific 
examples using Bethe ansatz in \ci{SS}. 
The equivalent general conclusion about 2-loop matching was 
obtained in the integrability-based approach in \ci{kmmz}. 
The order-by-order agreement seems  to break down 
at $\tl^3$ (3-loop) order, and a natural reason  
 \ci{SS,bds}
 is that the string limit (first $J \to \infty$, then 
 $\tilde \l \to 0$) and the SYM limit 
 (first $\l \to 0$, then  $J \to \infty$) need not be 
 the same. Suggestions how to  ``complete''
 the gauge-theory answer to have  the agreement 
 with string theory appeared in  \ci{bds,afs}.\foot{They 
 also resolve the order $\tl^3$ 
  disagreement \ci{cal} 
  between string and gauge theory 
  predictions for $1 \ov J$ corrections to the BMN spectrum.
  A possible explanation of why the agreement took place at
  first two orders in $\tl$ 
  is that the structure of the dilatation operator  at one and
  two loop order is, in a sense, 
  fixed by the BMN  limit, 
  which thus essentially determines the low energy effective
  action  in a unique way; this is no longer so starting with
  3-loop order.} 
  
 One can also generalize \ci{lopez,ST} 
  the above leading-order agreement to
 the $SU(3)$ sector of states with three large $S^5$ spins
 $J_i, 
 \ i=1,2,3$,  finding the  $CP^2$ analog of the 
 $CP^1$   ``Landau-Lifshitz''
  Lagrangian in \rf{con},\rf{coon} 
   \ci{ST} 
  $\LL = - i u^*_i \del_0 u_i - \ha \tl |D_1 u_i|^2 $
  on both string and spin chain sides. 
  Similar conclusion is reached  \ci{ST} in the $SL(2)$ sector of
  $(S,J)$  states (using the dilatation operator of \ci{BS}), 
  where  $\vn \to \vec l, \ l^2_1 - l^2_2 - l^2_3 =1$.
  Finally, one can consider also pulsating string states 
  discussed in the next section. 
  
\section{General fast motion in $S^5$ and scalar operators
from $SO(6)$ sector}

One would like to try to understand the general conditions on
string states  and SYM operators  for which the above 
correspondence works, and, in particular, incorporate also 
states with large oscillation numbers.
Here we will follow \ci{ST,kt} (a closely related approach was
developed in \ci{mik,mikk}).
For strings moving in $S^5$ with large oscillation  number 
$E= L + c_1 { \l \ov L} + ...$, i.e. the limit 
$L \to \infty, \ \tl= { \l \ov L^2}\to 0$ is again regular
\ci{Min1}, and the 
leading-order duality 
relation between string energies  and 
anomalous dimensions was  checked in \ci{bmsz,Min2,Min3}.
The general condition on string solutions  for which 
$E/L= f( \tl)$ has a regular expansion in $\tl$ 
appears to be that the world sheet metric should 
degenerate \ci{mik} in
the 
$\tl \to 0$ limit, i.e. the string motion should be  
ultra-relativistic 
in the small string tension limit \ci{mateos}. 
In the strict tensionless 
$\tl\to 0$ limit each string piece is following a
geodesic (big circle) of $S^5$, while switching on tension
leads to a  slight deviation from geodesic flow, i.e. to a 
nearly-null world surface \ci{mik}.  
The dual coherent SYM operators are then ``locally BPS'', 
i.e. each string bit corresponds to a BPS linear combination 
of 6 scalars. In general, the  scalar operators 
can be written as $$\O=  C_{m_1 ... m_L} \tr ( \p_{m_1} 
... \p_{m_L}) \ .$$ The planar 1-loop  dilatation operator 
acting on $ C_{m_1 ... m_L}$
 was found in \ci{mz1} (and is  equivalent 
to an integrable $SO(6)$ spin chain Hamiltonian)
\be
H_{m_1\cdots m_L}^{ n_1\ldots n_L} = \frac{\lambda}{(4\pi)^2} 
  \sum_{l=1}^{L} \left(\delta_{m_lm_{l+1}}\delta^{n_ln_{l+1}}+
2\delta^{n_l}_{m_l}\delta^{n_{l+1}}_{m_{l+1}}
-2\delta^{n_{l+1}}_{m_l}\delta^{n_{l}}_{m_{l+1}}\right)\ .
\label{Hamil}
\ee
To find the analog of the coherent-state action \rf{coon} 
 we choose a natural set of  coherent states 
 $\Pi_l \ket{v_l}$, where at each  site 
 $ \ket{v}= R(v) \ket{0}$. Here 
 $R$ is an $SO(6)$ rotation 
 and $\ket{0}$ is the BPS vacuum
 state corresponding to $\tr ( \p_1 + i \p_2)^L$ or 
 $v_{(0)} = (1,i,0,0,0,0)$, which is invariant 
 under $H= SO(2) \times SO(4)$. Then the rotation 
 $R(v)$  and thus the coherent state is parametrized by a point in   
 $G/H=SO(6)/[SO(4)\times SO(2)]$, i.e. $v$
  belongs to the
 Grassmanian $G_{2,6}$ \ci{ST}.
$G_{2,6}$ is  thus  the coherent state 
target space for the spin chain sigma model since it 
parametrizes the orbits of the  half-BPS operator
$\p_1+i\p_2$ under the $SO(6)$ rotations. 
 This is the space of 2-planes passing through zero 
  in $R^6$, or
  the space of big circles in $S^5$, 
 i.e. the moduli space of geodesics in $S^5$ 
 \ci{mikk}.  It can be represented also as an 8-dimensional
 quadric  in $CP^5$: a complex 6-vector $v_m$ 
 should be subject, in addition to $v_m v^*_m=1$
 (and gauging away the common phase) 
 also to  $v_m v_m=0$ condition. 
 Taking the limit $L\to \infty$ with fixed $\tl={\l \ov 
 L^2} $   and  the continuum limit $v_{lm}(t) \to 
 v_m(t,\s)$ we then get  the $G_{2,6}$  
 analog of the $CP^1$ action \rf{con},\rf{coon} 
 \be \la{ggg}
 S= L \int dt \inti \left( 
- i  v^*_m \del_0 v_m  -  \ha \tl | D_1 v_m|^2 \right) \ ,
 \ \ \ \ \ \  v_m v^*_m=1\ , \ \ \ \ v_m v_m=0 \ ,   \ee
 where 
 as in $CP^n$ case 
  $D_1 v_m = \del_1 v_m - (v^* \del_1 v) v_m $.
  
  One may wonder  how this 8-dimensional sigma model can be
  related to string sigma model on $R \times S^5$ where the
  coordinate space  of transverse motions is only 4-dimensional. 
  The crucial point is that the coherent state action is defined 
  on the phase space (cf. the harmonic oscillator case), 
  and $8=(1+5) \times 2 - 2 \times 2 $ is indeed the phase space 
  dimension of a string moving in $S^5$.
  On the string side, the need to use the phase space description is
  related to the fact that to isolate a ``fast'' coordinate $\a$ 
  for a generic string motion we need to specify both the position
  and velocity of each string piece. Given 
  $\LL= - (\del t)^2 + (\del X_m)^2 $ in conformal gauge 
  ($ \dot X X'=0, \ \dot X^2 + X'^2 =\k^2$, \ $X^2_m=1$) 
   we find the 
  geodesics as $X_m = a_m \cos \a + b_m \sin \a$, 
  where $\a =\k \tau, \ a^2_m=1, \ b^2_m=1, \ a_m b_m=0$.
  Equivalently, 
  $X_m = { 1 \ov \sqrt 2} ( e^{i\a } v_m + e^{-i\a } v^*_m)$,
   where $v_m = { 1 \ov \sqrt 2}(a_m - i b_m), \ 
   v_m v_m^*=1, \ v_m v_m=0$, i.e. the constant $v_m$ 
   belongs to $G_{2,6}$. In general, for near-relativistic
   string  motions 
    $v_m$ should  change slowly with $\tau$ and $\s$. 
   Then starting with the phase space Lagrangian for $(X_m, p_m)$ 
   $$L= p_m \dot X_m - \ha p_m p_m - \ha X'_m X'_m $$ 
  we may  change
   the variables according to \ci{kt}
   (cf. harmonic oscillator case)
\be X_m = { 1 \ov \sqrt 2} ( e^{i\a } v_m + e^{-i\a } v^*_m)\ , \ \ \ \ \ 
\ \ \  
   \ p_m = { i \ov \sqrt 2} p_\a ( e^{i\a } v_m - e^{-i\a } v^*_m)\ , \ee
   where $\a$ and $v_m$ 
 now depend on $\tau$ and $\s$
   and $v_m$ belongs to $G_{2,6}$. 
   There is an obvious $U(1)$ gauge invariance, $\a\to \a-\b, \ v_m
\to e^{i\b} v_m$. Gauge-fixing $t \sim \tau, \ p_\a \sim L$
(or, after approximate  T-duality in $\a$,\ 
 $\td \a \sim \s$)
one finds that the phase-space Lagrangian  becomes 
(after a rescaling of the time coordinate) 
\ci{kt}:
\be 
\LL\sim  p_\a D_0 \a - \ha \tl |D_1 v|^2 - {1 \ov 4}\tl  [ e^{2 i \a} 
( D_1 v)^2 + c.c.]\ . \ee
 The first term here produces $ -i v^*_m \del_0
v_m$  and the last term averages to zero  since 
$\a \approx \k \tau + ...$ where $\k= ({ \sqrt {\td \l}  })^{-1} 
 \to \infty$. 
Equivalently, the $\a$-dependent terms in the action 
(that were absent in the pure-rotation  $SU(3)$ sector) 
can be eliminated by canonical transformations \ci{kt}.
We then end up with the following 8-dimensional 
phase-space Lagrangian 
for the ``transverse'' string motions:
$ \LL = -i v^*_m \del_0 v_m  - \ha \tl | D_1 v_m|^2,$
which is the same as found on the spin chain side \rf{ggg}. 
The  3-spin $SU(3)$ rotation case is the special 
case when $v_m=(u_1,i u_1, u_2, i u_2, u_3, i u_3)$, 
where $u_i$ belongs to $CP^2$ subspace of $G_{2,6}$. 
The agreement between the spin chain and the 
string sides in this
general $G_{2,6}=
SO(6)/[SO(4)\times SO(2)]$ case explains not only 
the matching for pulsating solutions \ci{Min1,Min2} but also
 for near-by fluctuations.  
 
 
 Let us now 
  discuss the reason for the restriction $v^2=0$ 
 and also the structure of coherent operators  corresponding to
 semiclassical string states. 
 Given  $\O= C_{m_1\ldots m_L} \tr 
 \left(\p_{m_1}\ldots \p_{m_L} \right) $ one   
 may obtain the Schr\"odinger  equation for 
 the wave function $C(t)$ from\foot{For coherent states we consider 
 the equation 
 of motion that follows from this action 
 may be interpreted as a (non-trivial) 
   RG equation 
for the  coupling constant associated to  the operator 
$ \O$.}
\be 
S = -\int dt \bigg(  i 
C^*_{m_1\ldots m_L}
 \frac{d}{dt} C_{m_1\ldots m_L} 
  +
C^*_{m_1\ldots m_L} H_{m_1\cdots m_L}^{
 n_1\ldots n_L} C_{n_1\ldots n_L} \bigg) \ . \ee 
 In the limit $L\to \infty$  we may  consider
 the coherent state description  and assume the  factorized ansatz 
 $$C_{m_1\ldots m_L}= v_{m_1} ... v_{m_L} \ , $$ where each $v_{ml}$ 
 is a complex unit-norm vector \ci{kt}.
 The BPS case corresponding to totally symmetric traceless 
 $C_{m_1\ldots m_L}$ is represented by $v_{ml}= v_{m (0)}$, 
 $v_{(0)}^2 =0$.  Using \rf{Hamil} and substituting the 
 ansatz for $C$ into the above action one finds 
 \be \la{acta}
S = - \int dt \sum_{l=1}^{L} \bigg( i \bv_l \frac{d}{dt} v_l + 
   { \l \ov (4\pi)^2} 
\bigg[(\bv_l\bv_{l+1})(v_lv_{l+1}) + 2 
 - 2 (\bv_lv_{l+1})(v_l\bv_{l+1})\bigg]
 \bigg) \ . \ee
As expected \ci{mz1}, the Hamiltonian (second term) 
 vanishes  for the 
BPS  case when $v_l$ does not depend on $l$ and  $v^2=0$.
More generally, if we assume that $v_l$ is changing slowly
 with $l$ (\ie\ $v_l\simeq v_{l+1}$), 
then we find that \rf{acta} contains a  potential term 
$(\bv_l\bv_{l})(v_lv_{l})$ coming from the first ``trace'' structure in \rf{Hamil}.
This  term will lead  to large (order $\l L$ \ci{mz1}) 
shifts of anomalous dimensions, 
invalidating  low-energy  expansion, i.e. prohibiting one from taking the  
continuum limit  $
L\to \infty , \   \tl={\l\ov L^2}={\rm fixed}$, 
 and thus from establishing   correspondence with string theory 
along the lines of \ci{kru,krt,ST}. To 
get solutions
with  variations of $v_l$ from site to site  small we are 
to impose $
v^2_l=0 , \ 
l=1,\ldots,L $
 which minimizes the potential energy coming 
from the first term in  (\ref{Hamil}).
This  condition 
implies that the operator at { each} site is 
invariant under half of supersymmetries:
if $v^2=0$ the matrix $v_m\Gamma^m$ appearing in the 
 variation of the 
operator  $v_m \p_m $, i.e. $
 \delta_\epsilon (v_m  \p_m )= \frac{i}{2} 
 \bar{\epsilon} (v_m \Gamma^m) \psi $, 
 satisfies $(v_m\Gamma^m)^2=0$.
This means    that 
$v_m \p_m $ is invariant under the  variations 
associated with the null eigenvalues.
One  may thus call $v^2=0$   a ``{\it local BPS}'' condition
 since the preserved combinations of supercharges 
in general are different for each $v_l$, i.e.  the 
operator corresponding to  $C=v_1... v_L$ is not BPS.
Here ``local'' should be
 understood in the sense of the spin chain,  or,  
equivalently,  the spatial world-sheet direction.\foot{This 
generalizes the argument implicit in \ci{kru}; an 
equivalent proposal was made in \ci{mikk}. This is related to but different 
from the ``nearly BPS'' operators  discussed in \ci{mateos} 
(which,  by definition,   were those which become BPS in
 the limit $\l\to 0$).}
In the case when 
 $v_l$ are slowly changing   we
can take the continuum limit   as in \ci{kru,krt,ST}
by  introducing the 2-d field $v_m(t,\s)$ with 
$v_{ml} (t)= v_m(t, {2\pi l\ov L})$. Then  
 \rf{acta} reduces to \rf{ggg}
(all higher derivative terms are suppressed 
by powers of $1\ov L$ and the potential term 
is absent due to the condition $v^2=0$), i.e. 
\rf{acta} becomes 
  equivalent to the $G_{2,6}$  ``Landau-Lifshitz''
 sigma model which was derived from the phase space action  
 on the string side. \foot{The presence of the trace in the SYM operators 
 implies that 
we have to consider only spin chain states that are 
invariant under translations in $l$ or in $\sigma$.
 This  means that the momentum in the 
direction $\sigma$ should vanish:
 $P_\sigma=0$, or, equivalently, 
$ \int_0^{2\pi} \frac{d\sigma}{2\pi} 
\ v^*_m  \partial_\sigma v_m 
 =0$. 
This should be viewed as  a condition on the 
solutions $v_m(t,\s)$. The same condition appears 
on the string side 
from  a constraint.}

To  summarize, considering ultra-relativistic strings in $S^5$ one
can isolate a fast variable $\a$ (a ``polar angle'' 
in the string phase space) whose momentum $p_\a$ is large.
One may gauge fix $p_\a$ to be constant $\sim L$, or $\td \a \sim L
\s$, so that $\s$ or the ``operator direction'' 
on the SYM side 
gets interpretation  of  ``T-dual to fast coordinate'' direction. 
As a result, one finds 
 a local phase-space action with 8-dimensional 
target space (where one  no longer can 
eliminate 4 momenta 
without spoiling the locality). This action 
is equivalent to the Grassmanian $G_{2,6}$ 
Landau-Lifshitz  sigma model 
action appearing on the 
spin chain side.

We  thus get a precise mapping between 
 string solutions and 
 operators representing coherent spin chain states. 
 Explicit examples corresponding to pulsating 
and rotating solutions are given in \ci{kt}. 
In the continuum limit we may write  the operator 
 corresponding to the solution $v(t,\s)$ as 
$\O =  \tr ( \prod_\s  \vv(t,\s) )   ,\  
\vv \equiv  v_{m}(t,\s)  \p_{m}.$
This locally BPS coherent operator is the 
  SYM operator naturally associated to a
ultra-relativistic string solution.
The $t$-dependence of the string solution thus translates into the RG scale 
dependence of $\O$, while the $\s$-dependence describes
 the ordering of the factors under 
the trace. 

 In general,  semiclassical string states 
represented by classical string solutions 
should be dual to coherent spin chain states or
 coherent operators, 
which are different from the exact eigenstates 
of the dilatation operator 
but which should lead to the same energy or anomalous
 dimension  expressions. 
At the same time, the Bethe ansatz approach \ci{mz1,bmsz,bfst,kmmz}
is  determining the  exact eigenvalues 
of the dilatation operator.
The reason why the two approaches happen to be in agreement 
is that in the limit   we consider  the problem is 
essentially semiclassical, and because of  the integrability 
of the spin chain, its exact eigenvalues are not just 
well-approximated by the classical solutions
but are  actually exactly reproduced by them, i.e.  
(just as in the harmonic oscillator or 
 flat space string theory case)
the semiclassical coherent state  sigma model approach happens 
to be exact.

\section{Concluding remarks}

As  reviewed above, during the last year and a half
it was realized that 
there exists a remarkable  generalization  of
the  near-BPS (BMN) limit to  non-BPS  but ``locally-BPS'' 
sector of string/SYM states. This is   an important  progress 
in understanding of gauge-string duality at the
 quantitative level.
 The hope is to use this to 
 find   the  string/SYM spectrum 
 exactly, at least in a subsector of states. 
 The relation between phase-space action
 for ``slow'' variables  on the string side and the 
 coherent-state action on the SYM (spin chain) side 
 gives a very explicit picture of how string action 
 ``emerges'' from the gauge theory (dilatation operator).
 It implies not only the  equivalence between string energies 
 and SYM dimensions (established  to first two orders 
 in expansion in effective coupling $\tl$) 
 but also a direct relation between the string profiles and the
 structure of coherent SYM operators \ci{kru,kt}.

One may try  also  to   use the duality 
 as a tool  to uncover 
 the structure of planar SYM theory to all orders  in 
$\l$ by imposing the exact agreement with particular 
string solutions. 
For example, demanding the consistency with the BMN 
scaling limit (along with the superconformal algebra)  determines 
the structure of the full  3-loop SYM dilatation 
operator in the $SU(2)$ sector \ci{bks,beit}.
One can also use the BMN limit to fix only a 
 part of the dilatation 
operator but to all orders in $\l$ \ci{rt}.
Generalizing \rf{ferr},\rf{two}  and the 3- and 4-loop 
expressions in \ci{bks,beit} one  can organize \ci{SS,krt,rt}
the dilatation operator as an expansion in powers of 
$\Q_{k,l}= I - \vec \s_k \cdot \vec \s_l$ which 
reflect interactions between spin chain sites, 
$$D = \sum \Q  + \sum \Q\Q + \sum \Q\Q\Q + ... \ . $$ 
Here the products $\Q...\Q$ are ``irreducible'',
i.e. each site index appears only once. The $\Q^2$-terms first
appear at 3 loops, $\Q^3$-terms -- at 5 loops, etc. 
\ci{bks,beit}. 
Concentrating on the order $\Q$ part $D^{(1)}$
 of $D$ one can write (here $L=J$):
  \be D^{(1)}=\sum_{r=0}^\infty { \l^r \ov (4 \pi)^{r}}  
  \sum^{L}_{l=1} \De_{r}(l)\ , \ \ \ \ \ \ 
 \De_{r}(l) = \sum_{k=1}^r a_{r,k} \Q_{l,l+k}\ ,   \ee
 or 
 $
 D^{(1)}=   \sum_{l=1}^L
 \sum_{k=1}^{L-1}\  h_k (L,\l)\  \Q_{l,l+k} .$
 Demanding the agreement with the BMN limit 
  one can then determine the
 coefficients $a_{r,k}$ and thus the function 
  $h_k$ explicitly to all orders in $\l$ \ci{rt}. 
  In particular, for  large $L$, i.e. 
  when $D$ acts on  ``long'' operators, one finds 
  \be
\label{D}
D^{(1)} = \sum_{l=1}^L \sum_{k=1}^\infty
f_k(\lambda)\  \Q_{l,l+k} \ , \ \ \ \ \ \ \ \ \ \
f_k(\lambda)= \sum_{r=k}^\infty
{\lambda^r \over (4\pi)^{2r}} \ a_{r,l} \ , 
\ee
where  the coefficients $f_k(\lambda)$
can  be summed up in terms of
the standard Gauss hypergeometric function \ci{rt}
\be
\label{co}
f_k(\lambda) =
 \left( {\lambda \over 4 \pi^2}\right)^k
{\Gamma(k-\ha) \over 4\sqrt \pi \ \Gamma(k+1)}
\;\
{}_2 F_{1} (k-\ha, k+\ha; 2 k + 1; -
 {\lambda \over \pi^2})\ .
\ee
The function 
$f_k(\l) $ smoothly interpolates between the usual 
 perturbative expansion 
at small $\l$  and 
$f_k(\lambda)
\sim \sqrt{\lambda}  $
at strong $\l$ (which is the expected  behaviour 
of anomalous dimensions of ``long''  operators dual 
to  ``semiclassical''  states).  
Similar interpolating functions are expected to 
appear in anomalous dimensions of other SYM operators.
Also, 
$f_k$ goes  rapidly to zero   at large $k$, so we
get  a spin chain with short range interactions. 

One may hope that imposing additional constraints 
coming from correspondence with other string  solutions 
may help to determine the dilatation operator further
(see  also \ci{bds,afs}).

\section*{Acknowledgements}
We are grateful to 
 M. Kruczenski, 
 A. Ryzhov and  B. Stefanski 
for  collaboration on the work described above. 
This  work  was supported  by the  DOE
grant DE-FG02-91ER40690, the INTAS contract 03-51-6346
and the RS Wolfson award.


\begin{thebibliography}{00}

\bi{POL}
A.~M.~Polyakov,
``Gauge fields and space-time,''
Int.\ J.\ Mod.\ Phys.\ A {\bf 17S1}, 119 (2002)
[hep-th/0110196].

\bi {bmn}
D.~Berenstein, J.~M.~Maldacena and
H.~Nastase, ``Strings in flat space and pp waves
from N =4 super Yang Mills,''
JHEP {\bf 0204}, 013 (2002)
[hep-th/0202021].

\bi{mt}
R.~R.~Metsaev and A.~A.~Tseytlin,
``Type IIB superstring action in AdS(5) x S(5) background,''
Nucl.\ Phys.\ B {\bf 533}, 109 (1998)
[hep-th/9805028].



\bibitem{gkp}
S.~S.~Gubser, I.~R.~Klebanov and
A.~M.~Polyakov,
``A semi-classical limit of the
gauge/string correspondence,''
Nucl.\ Phys.\ B {\bf 636}, 99 (2002)
[hep-th/0204051].

\bi{mets}
R.~R.~Metsaev,
``Type IIB Green-Schwarz superstring in plane wave Ramond-Ramond  background,''
Nucl.\ Phys.\ B {\bf 625}, 70 (2002)
[hep-th/0112044].
R.~R.~Metsaev and A.~A.~Tseytlin,
``Exactly solvable model of superstring in plane wave Ramond-Ramond
background,''
Phys.\ Rev.\ D {\bf 65}, 126004 (2002)
[hep-th/0202109].



\bibitem{ft1}
S.~Frolov and A.~A.~Tseytlin,
``Semiclassical quantization of
rotating superstring in \adss,''
JHEP {\bf 0206}, 007 (2002)
[hep-th/0204226].



\bi{tse1}
A.~A.~Tseytlin,
``Semiclassical quantization of superstrings: \adss  and
beyond,''
Int.\ J.\ Mod.\ Phys.\ A {\bf 18}, 981 (2003)
[hep-th/0209116].


\bi{par}
A.~Parnachev and A.~V.~Ryzhov,
``Strings in the near plane wave background and AdS/CFT,''
JHEP {\bf 0210}, 066 (2002)
[hep-th/0208010].


\bi{cal}
C.~G.~Callan, H.~K.~Lee, T.~McLoughlin, J.~H.~Schwarz, I.~Swanson and X.~Wu,
``Quantizing string theory in \adss: Beyond the pp-wave,''
Nucl.\ Phys.\ B {\bf 673}, 3 (2003)
[hep-th/0307032].



\bi{grom}
D.~J.~Gross, A.~Mikhailov and R.~Roiban,
``Operators with large R charge in N = 4 Yang-Mills theory,''
Annals Phys.\  {\bf 301}, 31 (2002)
[hep-th/0205066].



\bibitem{bks}
N.~Beisert, C.~Kristjansen and M.~Staudacher,
``The dilatation operator of N = 4 super Yang-Mills
theory,''
Nucl.\ Phys.\ B {\bf 664}, 131 (2003)
[hep-th/0303060].

\bibitem{beit}
N.~Beisert,
``The su(2$|$3) dynamic spin chain,''
hep-th/0310252.

\bi{zan}
A.~Santambrogio and D.~Zanon,
``Exact anomalous dimensions of N = 4 Yang-Mills operators with large R
charge,''
Phys.\ Lett.\ B {\bf 545}, 425 (2002)
[hep-th/0206079].


\bi{kvs}
I.~R.~Klebanov, M.~Spradlin and A.~Volovich,
``New effects in gauge theory from pp-wave superstrings,''
Phys.\ Lett.\ B {\bf 548}, 111 (2002)
[hep-th/0206221].

\bi{SS}
D.~Serban and M.~Staudacher,
``Planar N = 4 gauge theory and the Inozemtsev long range spin chain,''
JHEP {\bf 0406}, 001 (2004)
[hep-th/0401057].

\bi{bds}
N.~Beisert, V.~Dippel and M.~Staudacher,
``A novel long range spin chain and planar N = 4 super Yang-Mills,''
hep-th/0405001.

\bi{afs}
G.~Arutyunov, S.~Frolov and M.~Staudacher,
``Bethe ansatz for quantum strings,''
hep-th/0406256.




\bi{gw}
D.~J.~Gross and F.~Wilczek,
``Asymptotically Free Gauge Theories. I,''
Phys.\ Rev.\ D {\bf 8}, 3633 (1973).


\bi{klv}
A.~V.~Kotikov, L.~N.~Lipatov and V.~N.~Velizhanin,
``Anomalous dimensions of Wilson operators in N = 4 SYM theory,''
Phys.\ Lett.\ B {\bf 557}, 114 (2003)
[hep-ph/0301021].


\bi{klvo}
A.~V.~Kotikov, L.~N.~Lipatov, A.~I.~Onishchenko and V.~N.~Velizhanin,
``Three-loop universal anomalous dimension of the Wilson operators in N = 4
SUSY Yang-Mills model,''
hep-th/0404092.

\bi{gkt}
S.~S.~Gubser, I.~R.~Klebanov and A.~A.~Tseytlin,
``Coupling constant dependence in the thermodynamics of N = 4  supersymmetric
Yang-Mills theory,''
Nucl.\ Phys.\ B {\bf 534}, 202 (1998)
[hep-th/9805156].



\bibitem{ft2}
S.~Frolov and A.~A.~Tseytlin,
``Multi-spin string solutions in
\adss,''
Nucl.\ Phys.\ B {\bf 668}, 77 (2003)
[hep-th/0304255].


\bibitem{ft3}
S.~Frolov and A.~A.~Tseytlin,
``Quantizing three-spin string
solution in \adss,''
JHEP {\bf 0307}, 016 (2003)
[hep-th/0306130].

\bibitem{ft4}
S.~Frolov and A.~A.~Tseytlin,
``Rotating string solutions: AdS/CFT duality in
non-supersymmetric
sectors,''
Phys.\ Lett.\ B {\bf 570}, 96 (2003)
[hep-th/0306143].


\bibitem{afrt}
G.~Arutyunov, S.~Frolov, J.~Russo and A.~A.~Tseytlin,
``Spinning strings in \adss and integrable systems,''
Nucl.\ Phys.\ B {\bf 671}, 3 (2003)
[hep-th/0307191].

\bibitem{art}
G.~Arutyunov, J.~Russo and A.~A.~Tseytlin,
``Spinning strings in \adss: New integrable system relations,''
Phys.\ Rev.\ D {\bf 69}, 086009 (2004)
[hep-th/0311004].

\bibitem{tse2}
A.~A.~Tseytlin,
``Spinning strings and AdS/CFT duality,''
hep-th/0311139.


\bibitem{mz1}
J.~A.~Minahan and K.~Zarembo,
``The Bethe-ansatz for N = 4 super
Yang-Mills,''
JHEP {\bf 0303}, 013 (2003)
[hep-th/0212208].

\bibitem{bmsz}
N.~Beisert, J.~A.~Minahan, M.~Staudacher and K.~Zarembo,
``Stringing spins and spinning strings,''
JHEP {\bf 0309}, 010 (2003)
[hep-th/0306139].

\bibitem{bfst}
N.~Beisert, S.~Frolov, M.~Staudacher and
A.~A.~Tseytlin,
``Precision spectroscopy of AdS/CFT,''
JHEP {\bf 0310}, 037 (2003)
[hep-th/0308117].



\bibitem{as}
G.~Arutyunov and M.~Staudacher,
``Matching higher conserved charges for strings and spins,''
JHEP {\bf 0403}, 004 (2004)
[hep-th/0310182].
``Two-loop commuting charges and the string / gauge duality,'' hep-th/0403077. 

\bibitem{Min2}
J.~Engquist, J.~A.~Minahan and K.~Zarembo,
``Yang-Mills duals for semiclassical strings on \adss,''
JHEP {\bf 0311}, 063 (2003)
[hep-th/0310188].

\bi{kru}
M.~Kruczenski,
``Spin chains and string theory,''
hep-th/0311203.


\bibitem{kmmz}
V.~A.~Kazakov, A.~Marshakov, J.~A.~Minahan and
K.~Zarembo,
``Classical/quantum integrability in
AdS/CFT,''
hep-th/0402207.


\bibitem{krt}
M.~Kruczenski, A.~V.~Ryzhov and A.~A.~Tseytlin,
``Large spin limit of \adss string theory and 
low energy expansion of
ferromagnetic spin chains,''
hep-th/0403120.

\bi{zarL}
M.~Lubcke and K.~Zarembo,
``Finite-size corrections to anomalous dimensions in N = 4 SYM theory,''
JHEP {\bf 0405}, 049 (2004)
[hep-th/0405055].

\bi{char}
C.~Kristjansen,
``Three-spin strings on \adss from N = 4 SYM,''
Phys.\ Lett.\ B {\bf 586}, 106 (2004)
[hep-th/0402033].
L.~Freyhult,
``Bethe ansatz and fluctuations in SU(3) Yang-Mills operators,''
hep-th/0405167.
C.~Kristjansen and T.~Mansson,
``The Circular, Elliptic Three Spin String from the SU(3) Spin Chain,''
hep-th/0406176.

\bi{BS}
N. Beisert and M. Staudacher, 
``The N=4 SYM integrable super spin chain'', 
Nucl.\ Phys.\ B {\bf 670}, 439 (2003)
[hep-th/0307042].

\bibitem{Min1}
J.~A.~Minahan,
``Circular semiclassical string solutions on \adss,''
Nucl.\ Phys.\ B {\bf 648}, 203 (2003)
[hep-th/0209047].


\bi{zhang}
W.~M.~Zhang, D.~H.~Feng and R.~Gilmore, 
``Coherent States: Theory And Some Applications,'' 
Rev.\ Mod.\ Phys.\ {\bf 62}, 867 (1990).

\bi{fra}
E.~H.~Fradkin,
``Field Theories Of Condensed Matter Systems,''
 Redwood City, USA: Addison-Wesley (1991) 350 p. (Frontiers in
physics, 82).
I. Affleck, ``Quantum spin chains and the Haldane gap'',
J. Phys C 1 (1989), 3047

\bibitem{lopez}
R.~Hernandez and E.~Lopez,
``The SU(3) spin chain sigma model and string theory,''
JHEP {\bf 0404}, 052 (2004)
[hep-th/0403139].

\bibitem{ST}
B.~J.~Stefanski, Jr.  and A.~A.~Tseytlin,
``Large spin limits of AdS/CFT and generalized Landau-Lifshitz equations,''
JHEP {\bf 0405}, 042 (2004)
[hep-th/0404133].



\bi{kt}
M.~Kruczenski and A.~A.~Tseytlin,
``Semiclassical relativistic strings in $S^5$ and long
 coherent operators in N =
4 SYM theory,''
hep-th/0406189.




\bibitem{mateos}
D.~Mateos, T.~Mateos and P.~K.~Townsend,
``Supersymmetry of tensionless rotating strings in \adss,
and nearly-BPS operators,''
hep-th/0309114.
``More on supersymmetric tensionless rotating strings in
\adss,''
hep-th/0401058.


\bibitem{mik}
A.~Mikhailov,
``Speeding strings,''
JHEP {\bf 0312}, 058 (2003)
[hep-th/0311019].

\bibitem{mikk}
A.~Mikhailov,
``Slow evolution of nearly-degenerate extremal surfaces,''
hep-th/0402067.
``Supersymmetric null-surfaces,''
hep-th/0404173.




\bibitem{Min3}
J.~A.~Minahan,
``Higher loops beyond the SU(2) sector,''
hep-th/0405243.

\bi{rt}
A.~V.~Ryzhov and A.~A.~Tseytlin,
``Towards the exact dilatation operator
 of N = 4 super Yang-Mills theory,''
hep-th/0404215.










\end{thebibliography}
\end{document}